\def\NAME#1#2{\caption{#2} \label{#1}}

\def\SEC#1#2{\section{#1} \label{#2}}
\def\SUB#1#2{\subsection{#1} \label{#2}}

\def\FIGT#1{\begin{figure}[!t] \begin{center} #1 \end{center} \end{figure}}
\def\FIGB#1{\begin{figure}[!b] \begin{center} #1 \end{center} \end{figure}}

\def\TABLET#1{\begin{table}[!t] \begin{center} #1 \end{center} \end{table}}

\def\TABLEWT#1{\begin{table*}[!t] \begin{center} #1 \end{center} \end{table*}}

\def\EPSH#1#2#3{\includegraphics[height= #2 cm,clip]{#1.eps} \NAME{#1}{#3}}
\def\EPSW#1#2#3{\includegraphics[width= #2 cm,clip]{#1.eps} \NAME{#1}{#3}}

\def\EPS08#1#2#3#4{\includegraphics[#3= #4 cm,clip]{../eps08/#1.eps} \NAME{#1}{#2}}

\def\AL#1{\begin{align} #1 \end{align}}

\def\EQN#1{\begin{eqnarray} #1 \end{eqnarray}}

\def\PM#1{ \begin{pmatrix} #1 \end{pmatrix} }

\newfont{\bg}{cmr10 scaled\magstep5}
\newfont{\bbg}{cmsy10 scaled\magstep5}

\newcommand{\bigzerou}{\smash{\lower2.7ex\hbox{\bg 0}}}

\def\T{\rule{0pt}{1.5ex}} 

\documentclass[twocolumn]{jsiamletters}
\usepackage{graphicx}
\usepackage{amsmath}
\usepackage{multirow}
\talkdate{2009/9/29}
\affiliation{Department of Aeronautics and Astronautics, School of Engineering, The University of Tokyo}{Tokyo 113-8656, Japan}
\affiliation{Research Fellow of the Japan Society for the Promotion of Science}{Tokyo 102-8471, Japan}
\affiliation{Meiji Institute for Advanced Study of Mathematical Sciences, Meiji University}{Kanagawa 214-8571, Japan}
\affiliation{Research Center for Advanced Science and Technology, The University of Tokyo}{Tokyo 153-8904, Japan}
\affiliation{School of Engineering Science, University of Science and Technology of China}{Hefei 230026, China}
\affiliation{PRESTO, Japan Science and Technology Agency}{Tokyo 102-0075, Japan}
\authorinfo{Daichi Yanagisawa}{1,2}{tt087068@mail.ecc.u-tokyo.ac.jp}
\authorinfo{Akiyasu Tomoeda}{3,4}{atom@isc.meiji.ac.jp}
\authorinfo{Rui Jiang}{5}{rjiang@ustc.edu.cn}
\authorinfo{Katsuhiro Nishinari}{4,6,1}{tknishi@mail.ecc.u-tokyo.ac.jp}
\title{Excluded Volume Effect in Queueing Theory}
\abstract{
We have introduced excluded volume effect, which is an important factor to model a realistic pedestrian queue, into queueing theory.
The probability distributions of pedestrian number and pedestrian waiting time in a queue have been calculated exactly.
Due to time needed to close up the queue, the mean number of pedestrians increases as pedestrian arrival probability ($\lambda$) and leaving probability ($\mu$) increase even if the ratio between them (i.e., $\rho=\lambda/\mu$) remains constant.
Furthermore, at a given $\rho$, the mean waiting time does not increase monotonically with the service time (which is inverse to $\mu$), a minimum could be reached instead.
}

\def\runauthor{Daichi Yanagisawa et al.}
\group{
Applied Integrable Systems
}
\begin{document}

\maketitle

\SEC{Introduction}{INTRO} Queueing theory \cite{queuebook1} is one
of the most famous and important theory in these days since it is
applied to many systems in the real world such as traffic systems
\cite{q&p0}, production systems \cite{q&p1}, networks \cite{q&i1},
and so on. The mathematical formulation for mean waiting time,
which is usually calculated by Little's theorem \cite{queuebook1},
is widely used due to its simplicity. In the queueing theory, the
state of a queue is represented by the number of jobs, which are
vehicles, pedestrians, and packets in networks. When there are
some jobs in the queue, one job is always receiving service, and
when it leaves the queue, the service for next job starts
immediately. This phenomenon is suitable for a queue of packets
since operation for next packet starts instantly by a computer.
However, it is not realistic for a queue of vehicles and
pedestrians since there is a delay of moving to service window due
to the excluded volume effect 
which is not included in the queueing theory.

The excluded volume effect is studied in detail by analyzing the
asymmetric simple exclusion process (ASEP) \cite{RB-ME-07}. Many
traffic models and pedestrian dynamics models are developed by
extending ASEP \cite{C&S&S,socialreview}. They are very successful
since the excluded volume effect works adequately to represent
real movement of vehicles and pedestrians. Therefore, we introduce
the excluded volume effect into the queueing theory for the first time
in this paper to develop a practical theory for a pedestrian
queue. The probability distributions and the means of both
pedestrian number and pedestrian waiting time in a queue are
calculated exactly and compared with those obtained from normal
queueing models.


\SEC{Models and Mathematical Analysis}{QM}

\SUB{Outline of the Three Queueing Models}{OTQM}

As a comparison, in addition to our excluded volume effect
introduced queue (E-Queue), we briefly review normal queue
(N-Queue) with continuous time (N-Queue (C)) (which is the most
famous queueing model known as M/M/1 \cite{queuebook1}), as well
as that with discrete time (N-Queue (P)). In E-Queue and N-Queue
(P), parallel update is adopted because it is realistic for one
dimensional pedestrian dynamics \cite{CR-AS-AS-TGF09}.
Fig. \ref{jsiaml01_nqeq} is a schematic view of N-Queue (P) and E-Queue.
At each time step a pedestrian arrives at the queue with
probability $\lambda$ and a pedestrian at the service window
(pedestrian A) leaves the queue with probability $\mu$. In N-Queue
(P) (Fig. \ref{jsiaml01_nqeq} (a)), pedestrian B moves to the
service window as soon as pedestrian A leaves the queue. In
contrast, pedestrian B cannot move to the service window at time
step $t+1$ in E-Queue (Fig. \ref{jsiaml01_nqeq} (b)) due to the
excluded volume effect, he/she moves there at time step $t+2$.

\SUB{Master Equations for N-Queue (P)}{NQ}

Since mathematical analysis on N-Queue (C) is described in detail in many books \cite{queuebook1}, we start from N-Queue (P).
The master equations in the stationary state are described as follows: 
\EQN{
P(0) &=& (1-\lambda ) P(0) + (1-\lambda ) \mu P(1), \\
P(1) &=& \lambda P(0) + (1-\lambda ) \mu P(2) \nonumber \\
&+& \{ \lambda \mu + (1-\lambda )(1-\mu ) \} P(1), \\
P(n) &=& \lambda (1-\mu ) P(n-1) + (1-\lambda ) \mu P(n+1) \nonumber \\
&+& \{ \lambda \mu + (1-\lambda )(1-\mu ) \} P(n) \hspace{0.3cm}
(n\geq 2), } where $P(n)$ represents the probability that there
are $n \in [0,\infty)$ pedestrians in the queue. 
Note that the stationary state exits only when $\lambda<\lambda_{cr}$ is satisfied. 
$\lambda_{cr}$ is a critical value of $\lambda$, and when $\lambda \geq \lambda_{cr}$, queue length tends to infinity.
By solving these recurrence equations among three terms, we obtain $P(n)$, and $P_W(t)$ $(t \in [0, \infty))$ (Probability distribution of the waiting time, i.e., time between a pedestrian arrives at the queue and leaves there.), $N=\sum_{n=0}^{\infty} nP(n)$ (Mean number of pedestrians in the queue), and $W=\sum_{t=0}^{\infty} tP_W(t)$ (Mean waiting time) are calculated as shown in Tab. \ref{pq_matome}.

\FIGT{\EPSH{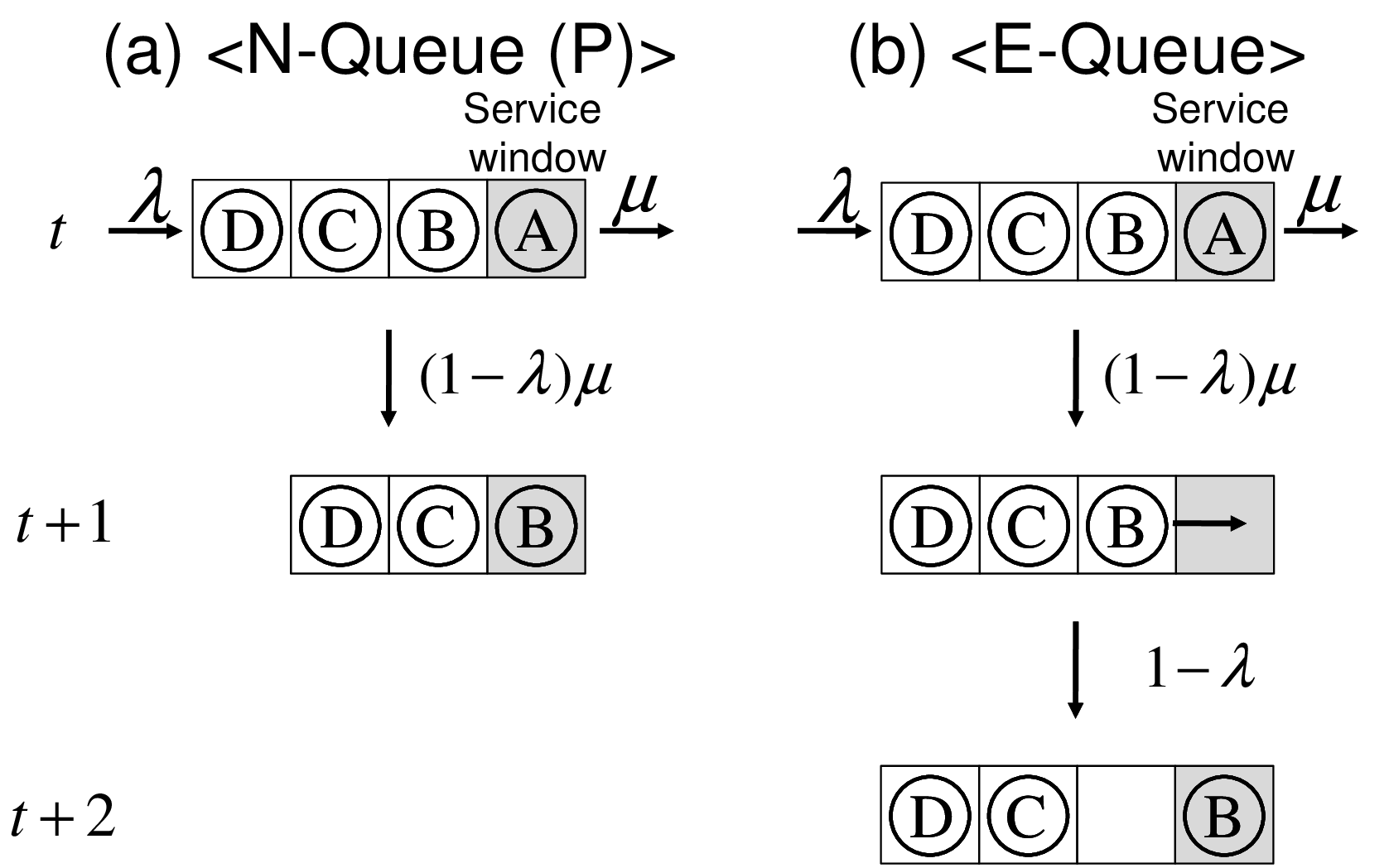}{5}{ Schematic views of time variation
of queueing states. (a) N-Queue (P). (b) E-Queue. The cell at the
right end in the queue is the service window. $\lambda \in [0,1]$
and $\mu \in [0,1]$ represent the arrival probability and the
service probability, respectively. }}

\FIGB{\EPSH{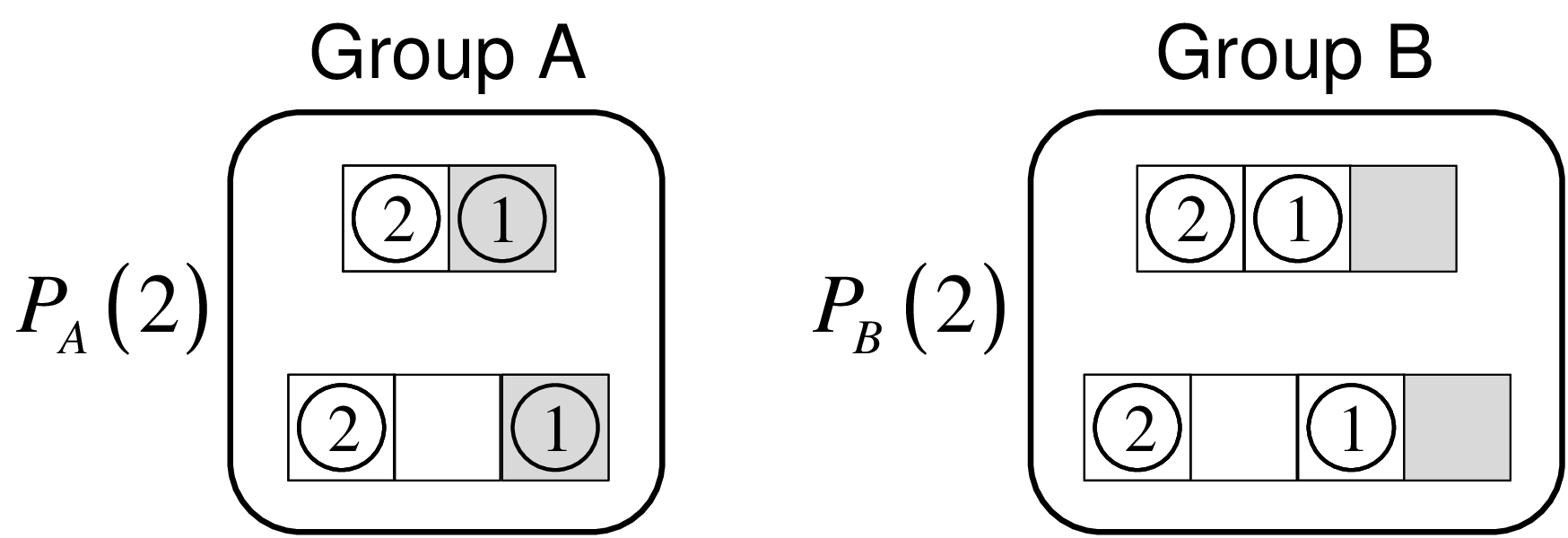}{2.7}{
Schematic views of the stationary states of E-Queue in the case $n=2$.
(a) Group A. The service window is occupied by a pedestrian.
(b) Group B. The service window is vacant.
}}

\FIGT{\EPSH{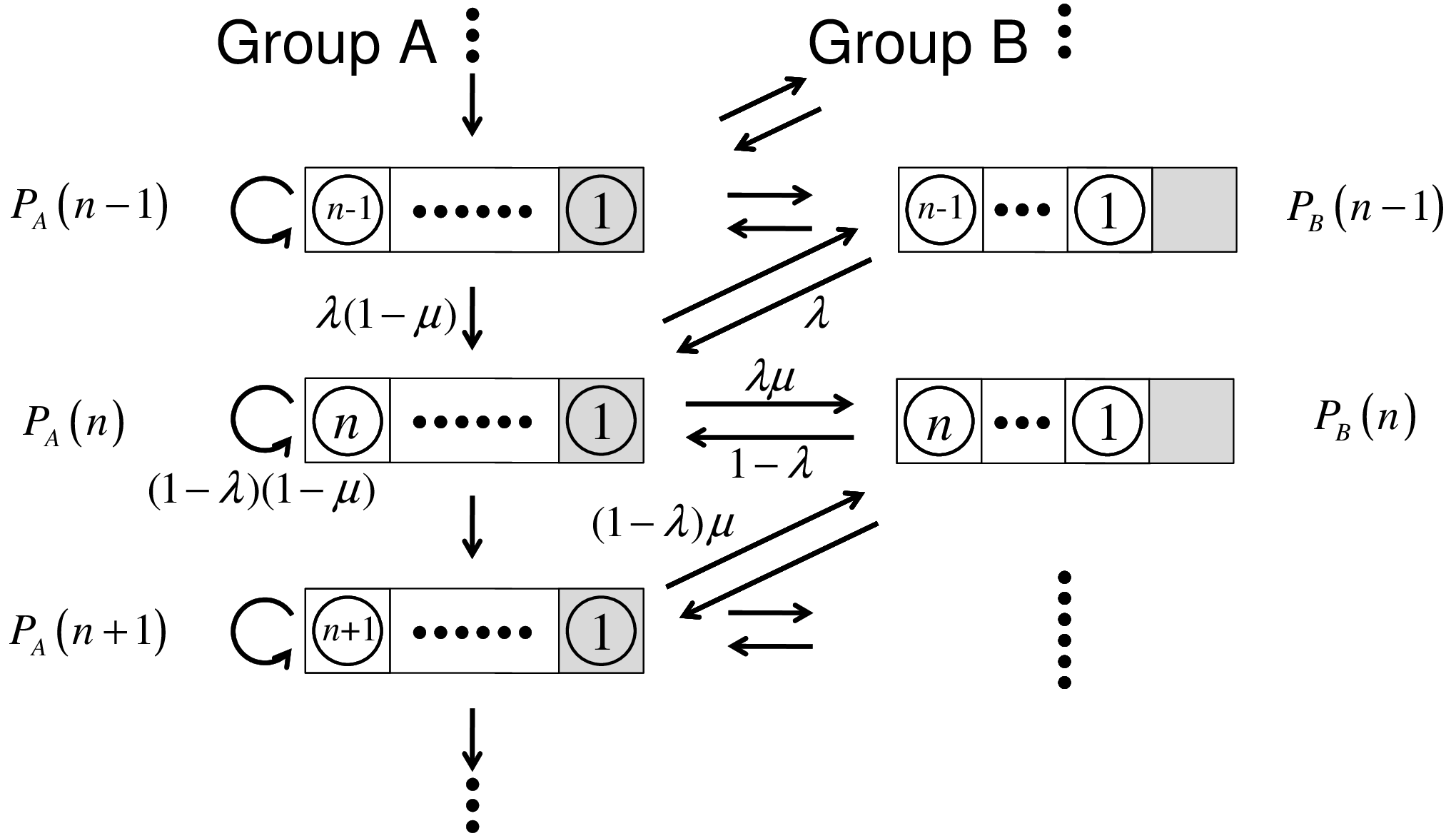}{4.7}{
State transition diagram of E-Queue.
}}

\SUB{Exact Solution for E-Queue}{EQ}

In E-Queue, the state is determined not only by pedestrian number
$n$. Fortunately, due to deterministic movement of pedestrians in
the queue (i.e., pedestrians move one cell in one time step if
their proceeding cell is vacant), two consecutive vacant cells
never appear in the stationary state. As a result, there are $2^n$
states when there are $n$ pedestrians in the queue since we only
need to consider whether there is a vacant cell or not in front of
each pedestrian. Schematic views of the stationary states in the
case $n=2$ are depicted in Fig. \ref{jsiaml02_onaji}.

Our target in this paper is to obtain probability distributions of
pedestrian number and pedestrian waiting time, so that the $2^n$
states do not need to be distinguished completely. The important
point is that whether the service window is occupied or not. Thus,
the $2^n$ states are divided into two groups A and B. The service
window is occupied in group A and it is vacant in group B. For
instance, two states belonging to group A, and the other two
states belonging to group B in the case $n=2$ as shown in Fig.
\ref{jsiaml02_onaji}.

\TABLEWT{ \NAME{pq_matome}{ Mathematical formulations of physical
quantities of three queueing models. The parameter $\rho=\lambda /
\mu$ represents the ratio between the mean service time ($1/\mu$) and
the mean arrival time ($1/\lambda$).
The expressions of $P(n)$ and $P_W(t)$ in the table are valid for
N-Queue (C) when $n\geq0$ and $t\geq0$, respectively, whereas,
those are valid in the two parallel update's queues only when
$n\geq1$ and $t\geq1$. $P_W(0)=0$ for N-Queue (P) and E-Queue.
Note that the Little's theorem $N = \lambda W$ is satisfied in all
three models. }
\begin{tabular}{cccc|c}
\hline \hline
Type        & N-Queue (C) & N-Queue (P) & E-Queue & Other Expressions for E-Queue \\
\hline \hline
$\lambda_{cr}$ & $\mu$ & $\mu$ & $\displaystyle \frac{\mu \T}{1+\mu}$ &
\multirow{2}{4cm}{ $\displaystyle P_A(n) = r^{n-1} \frac{\lambda}{(1-\lambda )\mu} P(0)$ } \\
\cline{1-4}
$P(0)$  & $1-\rho$  & $1-\rho$  & $\displaystyle 1-\frac{\rho \T}{1-\lambda}$ & \\
\cline{1-4}
$P(n)$  & $ (1-\rho) \rho^n $   &
$\displaystyle \frac{1-\rho}{1-\mu} \left( \frac{1-\mu}{1-\lambda} \rho \right)^n $ &
$\displaystyle \left( 1-\frac{\rho}{1-\lambda} \right) \frac{r^{n \T}}{1-\mu+\lambda \mu} $ &
\multirow{2}{4cm}{ $\displaystyle P_B(n) = r^{n-1} \frac{\lambda ^2}{(1-\lambda )^2 \mu} P(0)$ } \\
\cline{1-4}
$N$ & $\displaystyle \frac{\rho}{1-\rho}$   & $\displaystyle (1-\lambda) \frac{\rho}{1-\rho}$ & $\displaystyle \frac{\rho \T}{1-\frac{\rho}{1-\lambda}} $ & \\
\cline{1-4} \cline{1-4}
$P_W(t)$ & $\mu (1-\rho) \exp[-\mu(1-\rho)t]$ &
$\displaystyle \frac{\mu (1-\rho)}{1-\lambda}  \left( \frac{1-\mu}{1-\lambda} \right)^{t-1}$ &
$\displaystyle \left( \mu - \frac{\lambda}{1-\lambda} \right) \left( \frac{1}{1-\lambda} - \mu \right)^{t-1 \T}$ &
\multirow{2}{4cm}{ $\displaystyle r = \frac{ 1 - \mu + \lambda \mu }{(1-\lambda )^2 } \rho $} \\
\cline{1-4}
$W$ & $\displaystyle \frac{1^{\T}}{\lambda} \frac{\rho}{1-\rho}$ &
$\displaystyle \frac{1-\lambda }{\lambda} \frac{\rho}{1-\rho}$ &
$\displaystyle \frac{ \rho }{ \lambda \left( 1 - \frac{\rho}{1-\lambda} \right) } $ & \\
\hline
\end{tabular}
}

We describe the sum of the probabilities of the stationary states
in group A as $P_A(n)$ and that in group B as $P_B(n)$ when there
are $n$ pedestrians in the queue. Thus, $P(n)= P_A(n) + P_B(n)$.
Note that 
we have $P_A(0)=0$ and $P_B(0)=P(0)$.
The state transition diagram of E-Queue is depicted as Fig.
\ref{jsiaml03_seni} and the master equations in the stationary
state are described as follows:
\EQN{
P_A(1) &=& (1-\lambda )(1-\mu ) P_A(1) + \lambda P_B(0) \nonumber \\
&+& (1-\lambda ) P_B(1), \\
P_A(n) &=& \lambda (1-\mu ) P_A(n-1) \nonumber \\
&+& (1-\lambda )(1-\mu ) P_A(n) + \lambda P_B(n-1) \nonumber \\
&+& (1-\lambda ) P_B(n) \hspace{2cm} (n\geq2), \\
P_B(0) &=& (1-\lambda ) P_B(0) + (1-\lambda ) \mu P_A(1), \\
P_B(n) &=& \lambda \mu P_A(n) + (1-\lambda )\mu P_A(n+1) \nonumber \\
&\ & \hspace{4cm} (n \geq 1).
}
These equations could also be obtained by reducing the master equations where all $2^n$
states are distinguished \cite{dy_preparation}.
Solving the equations with normalization condition $\sum_{n=0}^{\infty} P(n) = 1$, we obtain the solutions in Tab. \ref{pq_matome} in the case $\lambda < \lambda_{cr}$.
The probability distribution of the waiting time $P_{W}(t)$ is also calculated as
\AL{
P_W(t) =
&f(t,1)P(0) + \sum_{n=1}^{Q(t,2)} [ f(t-n,n) \mu P_A(n) ] + \hspace{1.5cm} \nonumber \\
\sum_{n=1}^{Q(t-1,2)} &\Big[ f(t-n,n+1) \{ (1-\mu )P_A(n) + P_B(n) \} \Big]
} where $Q(a,b)$ returns a quotient of $a/b$, and $ f(t,n) = \PM{
t-1 \\ n-1 } \mu ^n (1-\mu )^{t-n}
$
is the negative binomial distribution. 
The means, i.e., $N$ and $W$, are also obtained and described in Tab. \ref{pq_matome}.

\SEC{N-Queue V.S. E-Queue}{COMP}

In this section, we compare physical quantities of the three queueing models. 
Note that $\lambda$ and $\mu$ are probabilities
in N-Queue (P) and E-Queue, while they are rates in N-Queue (C).
Besides, $t$ is discrete in the former models, whereas it is
continuous in the latter one. However, if we regard one time step
in the former two models as unit time in N-Queue (C), we can
compare the physical quantities of the models in the same time scale. 
The physical quantities are described by three parameters, which are $\lambda$, $\mu$, and $\rho$ as in Tab. \ref{pq_matome}.
However, since $\rho = \lambda / \mu$, there are only two independent variables; thus, we use $\rho$ and $\mu$ as independent ones in the following.
Then, $\lambda$ becomes a function of $\rho$ and $\mu$ described as $\lambda(\rho,\mu) = \rho \mu$.

\SUB{Critical Value $\lambda_{cr}$}{CL}
In N-Queues, the critical value $\lambda_{cr}=\mu$.
In E-Queue, $\lambda_{cr}=\mu / (1+\mu)$.
Since
$
\mu / (1+\mu) \leq \mu
$
, the region where stationary state exists in $\rho-\mu$ space is smaller in E-Queue than in N-Queues.
Due to time needed to close up vacant cells, which equals to one time step, the length of E-Queue diverges easier than that of N-Queues.

\SUB{Mean Pedestrian Number in a Queue}{MNPQ}

This subsection focuses on mean pedestrian number $N$ in a queue, which are denoted as $N_1$, $N_2$ and $N_3$ in N-Queue (C), N-Queue (P), and E-Queue, respectively.
As can be seen from Tab. \ref{pq_matome} and Fig. \ref{jsiaml04_gNP} (a),

(i) At given $\mu$, $N_1$, $N_2$ and $N_3$ increase monotonically with the increase of $\rho$, which is defined as $\rho=\lambda/\mu$.

(ii) At given $\rho$, $N_1=N_2=N_3$ when $\mu \rightarrow 0$
(which implies $\lambda \rightarrow 0$). With the increase of
$\mu$, $N_1$ remains unchanged, $N_2$ decreases and $N_3$
increases.

\FIGB{\EPSW{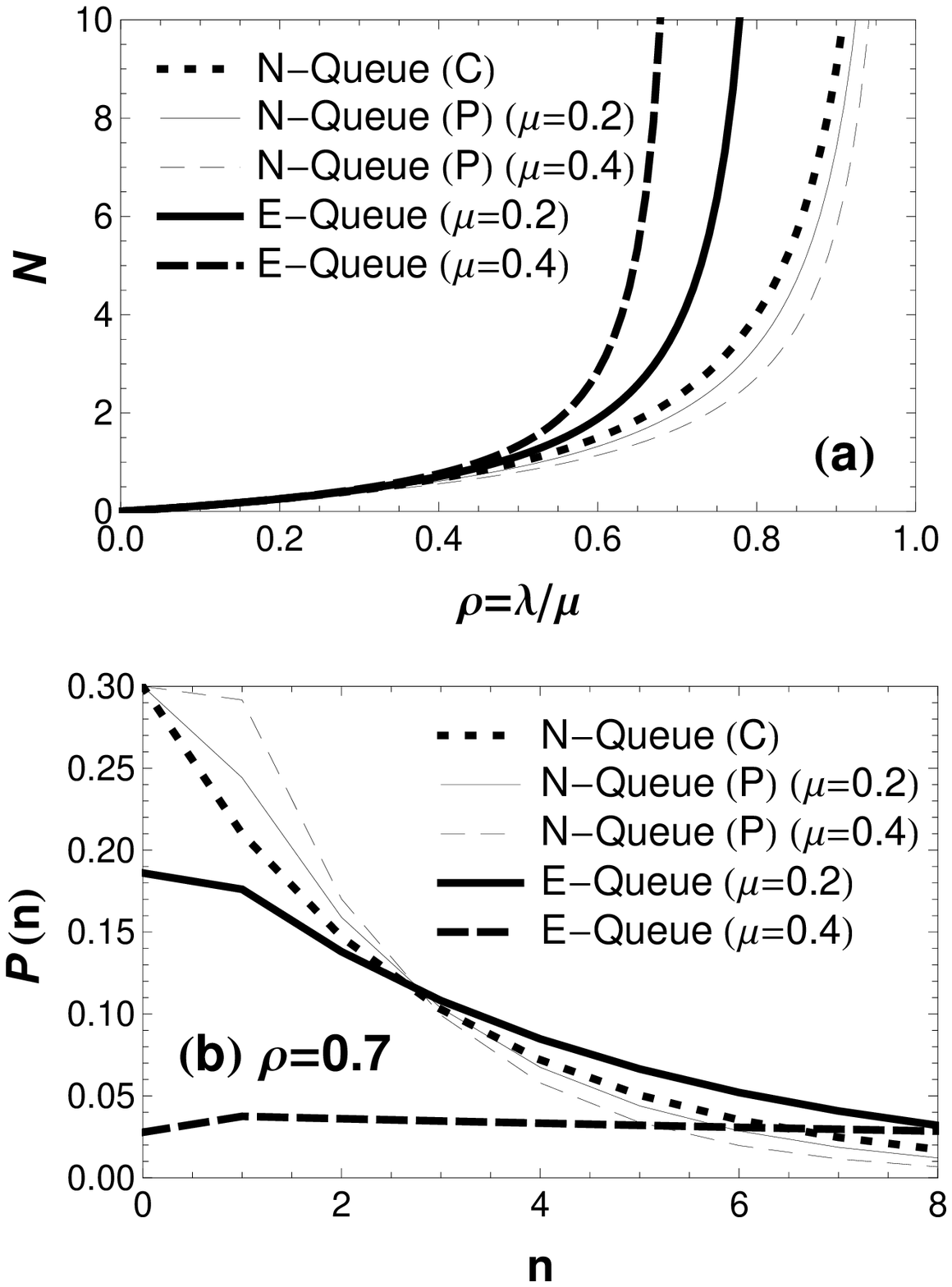}{9}{ (a) Mean number of pedestrians $N$
against the ratio between service and arrival time $\rho$. (b)
Probability distributions of pedestrian number in a queue. }}

In N-Queue (C), in an infinitesimal time interval $\Delta t$, the arrival probability and the leaving probability are $\lambda\Delta t$ and $\mu\Delta t$, which tends to zero. 
Thus, changing $\lambda$ and $\mu$ implies a rescaling of time interval, which does not change probability distribution of pedestrian number.
Therefore, $N_1$ is independent of $\mu$.
In N-Queue (P), $P(0)$ is independent of $\mu$; however, since the common ratio in $P(n)$ $\left( \frac{1-\mu}{1-\rho \mu} \rho \right)$ decreases with the increase of $\mu$, $P(n)$ becomes narrower and higher as in Fig. \ref{jsiaml04_gNP} (b).
Consequently, $N_2$ decreases as $\mu$ increases. 
In E-Queue, with the increase of $\mu$, $P(0)$ dramatically decreases as in Fig. \ref{jsiaml04_gNP} (b) and the common ratio $r$ increases when $\rho > 1/(2+\mu)$, so that $P(n)$ becomes wider and lower; thus, $N_3$ increases. 
We would also like to explain this phenomenon intuitively.
A pedestrian takes ``the time of
closing up" plus ``the service time", denoted as an ``extended
service time", to go through the service window. When $\mu$
increases the extended service time does not sufficiently decrease
since the time of closing up remains as a constant. At given
$\rho$, the increase of $\mu$ implies the increase of $\lambda$,
hence, $N_{3}$ increases as $\mu$ increases.

\FIGT{\EPSW{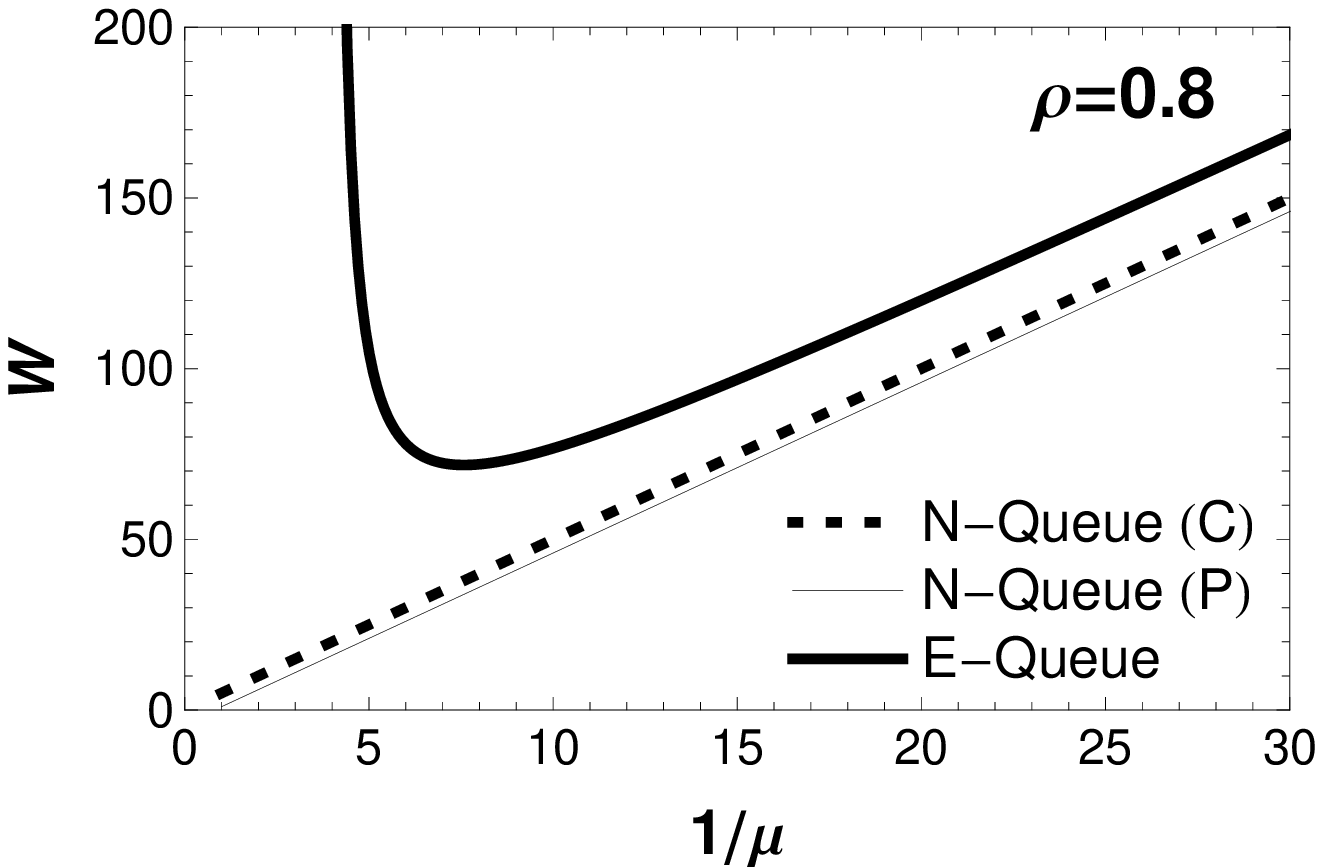}{9}{
Mean waiting time $W$ against the mean service time $1/\mu$.
}}

\SUB{Mean Waiting Time in a Queue}{MWTQ}

Figure \ref{jsiaml04_gW} shows the variation of $W$
against $1/\mu$, which is a mean service time, in the case $\rho$
is constant. With the increase of $1/\mu$, $W$ increases linearly
in N-Queue (C) and quasi-linearly in N-Queue (P), which coincides
with our intuition. $W$ also increases quasi-linearly in E-Queue
when $1/\mu$ is large; however, when $1/\mu$ is small, it
surprisingly achieves minimum $W_{\text{min}}(= \rho /
(1-\sqrt{\rho})^2)$ at $1/\mu_{\text{min}}$ $(
= \rho / (1-\sqrt{\rho}) )$, increases as $1/\mu$ further
decreases, and diverges at $1/\mu_{\text{cr}}$ $(
= \rho / (1-\rho) )$. Since the Little's theorem $N=\lambda W$
\cite{queuebook1} is satisfied in all three models,
$W=(1/\mu)(N/\rho)$. At a given $\rho$, with the increase of
$1/\mu$, $N$ does not change in N-Queue (C) and increases in
N-Queue (P), thus $W$ increases in both cases. In contrast,  $N$
decreases with the increase of $1/\mu$ in E-Queue, hence, the
minimum could be reached.

As we have seen in Fig \ref{jsiaml04_gW}, E-Queue becomes similar
to N-Queues when service time is large and different from it when
service time is small. Thus, it is useful to know quantitatively
when we should consider the excluded volume effect from the
perspective of application. In Fig. \ref{jsiaml05_gG}, the
$\rho$-$\mu$ plane is divided by the curve $R=1.1$, where $R$ is a
ratio between $W$ in N-Queue (C) and E-Queue described as
$R(\rho,\mu) = W_{\text{E-Queue}} / W_{\text{N-Queue (C)}}$
In the lower-left region in Fig. \ref{jsiaml05_gG} $R<1.1$, and the difference of $W$ is not critically large, so that it may be allowed to use N-Queue (C) for simple calculation when both $\rho$ and $\mu$ are small.
In contrast, $R>1.1$ and the difference is crucial in the upper-right region, therefore, the excluded volume effect should be considered when both $\rho$ and $\mu$ are large.
Note that Fig. \ref{jsiaml05_gG} is an example of the dividing curve, and it is possible to depict the other curves by determining $R$ as a different value.
Thus, this diagram is helpful to know the error quantitatively and make a decision whether to use N-Queue or E-Queue for designing a queueing system for pedestrians.

\FIGT{\EPSH{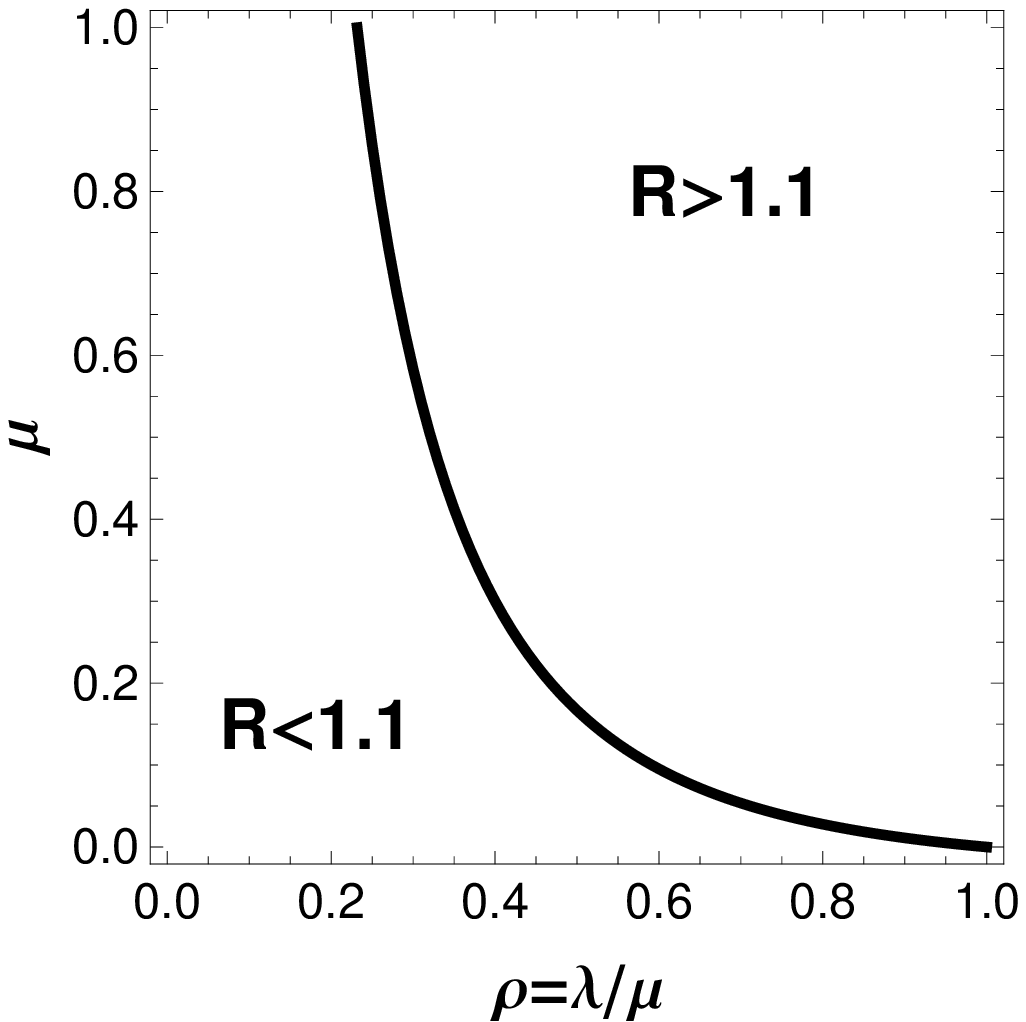}{6.1
}{
Curve of $R=1.1$ on the $\rho-\mu$ plane.
}}

\SEC{Conclusion}{CONC}

In this paper, we have newly introduced the excluded volume effect
into the queueing model and obtained the probability distributions and
the means of the number of pedestrians and the waiting time exactly. 
Such physical quantities of the new model are
compared with those of previous models which do not include the
excluded volume effect. When the service time is large enough, all
models become similar; however, when the service time becomes
small, the time of closing up the queue dominates and the waiting
time surprisingly increases in the excluded volume effect
introduced queueing model. 
We also obtain the diagram to examine when the effect of the excluded volume effect becomes prominent.

To construct a more realistic model, the movement in the queue should be stochastic since pedestrians do not homogenously close up the queue.
Furthermore, the length of a queue should be also calculated exactly, and the validity of the model should be verified by real experiments or observations in the near future.

Finally, we would like to mention that this work is financially supported by JSPS and JST.

\references

\end{document}